\documentclass[lineno]{biometrika}

\usepackage{setspace}
\usepackage{amssymb,caption,fancyhdr}
\usepackage{graphicx,amscd,amsmath,amssymb,amsfonts,verbatim}
\usepackage{mathrsfs,graphics}
\usepackage{bbm}
\usepackage[all]{xy}
\usepackage{xcolor}
\usepackage{bm}
\usepackage{bbm}

\usepackage{graphics,graphicx}
\usepackage{bbm}
\usepackage{amsmath}
\usepackage{amssymb}
\usepackage[all]{xy}

\usepackage{verbatim}
\usepackage{mathrsfs}
\usepackage{bbm}

\DeclareMathAlphabet{\mathpzc}{OT1}{pzc}{m}{it}

\def\E{\mathbb{E}}
\def\E{\mathbbmss{E}}
\def\P{\mathbbmss{P}}

\def\bY{\mathbf{Y}}

\def\bX{\mathbf{X}}

\renewcommand\hat{\widehat}

\renewcommand{\qed}{\nobreak \ifvmode \relax \else
      \ifdim\lastskip<1.5em \hskip-\lastskip
      \hskip1.5em plus0em minus0.5em \fi \nobreak
      \vrule height0.75em width0.5em depth0.25em\fi}

\usepackage{mathptmx}

\usepackage{graphics}

\begin{document}

\title{Selection bias when using instrumental variable methods to compare two treatments but more than two treatments are available}

\author{Ashkan Ertefaie$^{1}$, Dylan Small$^{2}$, James Flory$^3$ and \\ Sean Hennessy$^4$ \\
\textit{  $^{1}$Department of Statistics,  University of Pennsylvania; Center for Pharmacoepidemiology Research and Training, University of Pennsylvania.\\
$^{2}$Department of Statistics, University of Pennsylvania\\
$^{3}$ Weill-Cornell School of Medicine\\
$^{4}$Center for Pharmacoepidemiology Research and Training, Center for Clinical Epidemiology and Biostatistics, Perelman School of Medicine, University of Pennsylvania } }

\maketitle 

\begin{abstract}
Instrumental variable (IV) methods are widely used to adjust for the bias in estimating treatment effects caused by unmeasured confounders in observational studies. In this manuscript, we provide empirical and theoretical evidence that the IV methods may result in biased treatment effects if applied on a data set in which subjects are preselected based on their received treatments. We frame this as a selection bias problem and propose a procedure that identifies the treatment effect of interest as a function of a vector of sensitivity parameters. We also list assumptions under which analyzing the preselected data does not lead to a biased treatment effect estimate. The performance of the proposed method is examined using simulation studies. We applied our method on The Health Improvement Network (THIN) database to estimate the comparative effect of metformin and sulfonylureas on weight gain among diabetic patients. 
\end{abstract}


\section{Introduction} \label{intro}

Confounder adjustment is critical in the estimation of treatment effect in observational studies. In practice, however, there is no guarantee that all the confounders (i.e., baseline variables that affect both the treatment and outcome) have been measured.  These unmeasured confounders can  bias the treatment effect estimation. 

Instrumental variable (IV) methods are designed to adjust for the bias caused by unmeasured confounders in estimating a difference measure of effect such as the risk difference. IVs are predictors of treatment allocation  that are independent of unmeasured confounders and do not have a direct effect on the outcome (i.e., only affect the outcome through affecting the treatment).  Intuitively, the IV method seeks to extract variation in treatment that is free of unmeasured confounders  and uses this variation to estimate the treatment effect. For more information on IV methods, see \cite{angrist1996identification}, \cite{newhouse1998econometrics}, \cite{greenland2000introduction}, \cite{hernan2006instruments}, \cite{cheng2009semiparametric}, \cite{baiocchiatutorial} and \cite{imbens2014instrumental}.   This paper is motivated by provider preference IV (PP IV) which is commonly used as an IV in health studies. PP IV utilizes the natural variation in medical  practices to construct an IV.  For example, in studying the effect of cyclooxygenase-2 (COX-2) inhibitors vs. nonselective nonsteroidal anti-inflammatory drugs (nonselective NSAIDs) on gastrointestinal complications \cite{brookhart2006evaluating} used whether a patient's physician last prescription for a patient prescribed a nonsteroidal anti-inflammatory drug (NSAID) was for a Cox-2 inhibitor or a nonselective NSAID as an IV for whether the patient was prescribed a Cox-2 inhibitor or a nonselective NSAID.  Other examples of PP IV can be found in  \cite{brooks2003breast}  and \cite{brookhart2007preference}.

For a binary IV and binary treatment, \cite{angrist1996identification} proposed a nonparametric  estimator  which identifies the treatment effect among subjects who would take the treatment if encouraged to do so by the IV and not take the treatment if not encouraged to do so by the IV (compliers).  This estimator is equivalent to a  2 stage least square (2SLS) estimator where at the first stage the treatment is regressed on IV and at the second stage the outcome is regressed on fitted values of the first stage regression (i.e., predicted value of the treatment variable given the IV).  The coefficient of the independent variable of the second stage regression is equivalent to the estimator proposed by \cite{angrist1996identification}.  The 2SLS estimator is also called the Wald estimator after \cite{wald1940fitting}. More efficient versions of the 2SLS estimator has been developed by \cite{imbens1997estimating}, \cite{imbens1997bayesian} and \cite{cheng2009efficient}. 

Although treatment effect estimation using IV approaches  for binary treatments is straightforward, it is challenging  when there are more than two treatment arms. In fact, \cite{cheng2006bounds} shows that the treatment effects are not point identifiable and proposes bounds on treatment effects using method of moments. \cite{long2010estimating} studies the same problem and proposes a Bayesian approach for finding credibility intervals for the identification  region.

Sometimes  investigators are interested in comparing the effect of two treatments while the data consist of more than two treatment options for patients. In this situation a common approach is to  select patients who are assigned to the treatments of interest and perform the standard IV approach for binary treatment and IV to estimate the treatment effect \citep{hadley2003exploratory,  basu2007use, hadley2010comparative, suh2012comparative, chen2014comparative}.


The objective of this manuscript is to provide empirical and theoretical evidence that, in general, excluding patients based on their assigned treatment can result in selection bias. This problem has also been discussed in an independent work by \cite{swanson47}, \cite{ swansonphd}, and \cite{ swanson2015selecting}. In the current manuscript, we formalize the mechanism of this selection bias within a principal stratification framework and provide a procedure that identifies the treatment effect of interest as a function of a vector of sensitivity parameters. Specifically, the sensitivity parameters are used to estimate the probability of being assigned to one of the treatments of interest as a function of the observed outcome and the IV value \citep{gilbert2003sensitivity}. We also list assumptions under which selecting patients based on their received treatment does not lead to a biased treatment effect estimate.



The manuscript is organized as follows. Section \ref{sec:notation} provides the setup and defines the causal estimand. It also lists the required assumptions for our procedure.  In Section \ref{sec:ivbias}, we discuss the selection bias issue and present the proposed sensitivity analysis method. We report on the results of a simulation study in Section \ref{sec:simulation}, where we examine the performance of our proposed method. In Section \ref{sec:application}, we study the comparative effect of metformin and sulfonylureas on weight gain using The Health Improvement Network (THIN) database. 

\section{Settings and Notations} \label{sec:notation}

For simplicity we focus on  observational studies with three treatment options A, B, and C.  We assume that we are only interested in comparing the effect of treatment A with B. We use small letters to refer to the possible values of the corresponding capital letter random variable.  Let $Z$ be a binary instrumental variable where  $Z\in\{A,B\}$ such as provider preference.  $Z=A$ means that the IV predisposes  the patient to receive treatment level $A$ among the options $A$ and $B$, and $Z=B$ means that the  IV predisposes the patient to receive treatment level $B$ among the options $A$ and $B$; e.g., in \cite{brookhart2006evaluating} study, if A denotes Cox-2 inhibitor and $B$ denotes nonselective NSAID, $Z=A$ if the patient's physician last prescription to a patient prescribed an NSAID was a Cox-2 inhibotor and $Z=B$ if it was a nonselective NSAID.  Let  $D(z)$ be the potential treatment assigned  given $Z=z$, where  $D(z)\in\{A,B,C\}$, i.e., $D(A)$ is the treatment that would be received if the IV was $A$.  The potential outcomes are $Y^{(z,d)}$ for $Z=z$ and $D=d$, i.e., $Y^{(A,B)}$ is the outcome that would be observed if the IV was $A$ and the treatment level was $B$. The observed treatment and outcome are $D=D^Z$ and $Y= Y^{Z,D^Z}$, respectively. 

\cite{frangakis2002principal} developed a basic principal   strata (PS)  framework which stratifies subjects based on their potential assigned treatments. In our setting, we can classify subjects into the following 9 principal   strata: $S1$) always B taker, $S2$) always A taker, $S3$) always C taker, $S4$) subjects who would comply with their value of instrumental variable, i.e., $D(A)=A$, $D(B)=B$, $S5$) subjects who would take B if $Z=B$ and would take C if $Z=A$, $S6$)  subjects who would take C if $Z=B$ and would take A if $Z=A$, $S7$) subjects who would take A if $Z=B$ and would take B if $Z=A$, $S8$) subjects who would take B if $Z=A$ and would take C if $Z=B$, and $S9$) subjects who would take A if $Z=B$ and would take C if $Z=A$. We denote the proportions in the basic principal strata as $\pi^{S.}$.

We assume that the IV satisfies the following standard  assumptions \citep{angrist1996identification}: 1) stable unit treatment value assumption, 2) The IV is independent of unmeasured confounders, potential outcomes and potential treatment received, 3) the IV affects the treatment but has no direct effect on the outcome. This assumption is  known to as the exclusion restriction (ER) assumption.    Under the exclusion restriction, we let $Y^d$  be the potential outcome if treatment $d$ is received which equal $Y^{z=0,d}=Y^{z=1,d}$,    and 4) the IV is independent of the pretreatment covariates, potential outcome and potential treatments. Furthermore, we assume two types of monotonicity assumptions: 5) monotonicity I: there is no defiers (i.e., $p(D(A)\ \neq A,D(B)\neq B)=0$) and 6) monotonicity II: there is no one who would take $B$ if $Z=A$ but take $C$ if $Z=B$ and no one who would take $C$ if $Z=A$ but take $A$ if $Z=B$. Note that monotonicity assumptions 5 \& 6 reduce the number of principal   strata to 6. The parameter of interest is defined as 
\[
\theta = \E[Y^A-Y^B | PS=S4].
\] 


\section{IV Selection Bias} \label{sec:ivbias}
We show in this manuscript that IV analysis methods may not result in an unbiased treatment effect if applied on a data set in which units are preselected based on their received treatments.   A standard IV analysis of patients on the subset of patients for whom $D=A$ or $B$ is essentially a comparison of the $Z=A$ vs. $Z=B$ patients on this subset.  If $Z=A$ has a different effect on whether $D=C$ than $Z=B$, then there will be selection bias.  For example, patients who are in a more severe stages of the disease may be treated with treatment $D = C$ and therefore have less chance to be observed in the sample; if doctors who prefer drug $A$ $(Z=A)$ are more likely to use treatment $C$ with severe patients than doctors who prefer drug $B$ $(Z=B)$, then there will be selection bias.  Another scenario where such selection bias might arise is that treatment $C$ is more likely to be offered if some complications are present and doctors who prefer drug $A$ are more likely to treat a patient with complications with drug $C$ than doctors who prefer drug $B$.


The challenge is that $\theta$ is not identifiable using the observed data when there is selection bias.  This is due to the fact that the observed outcomes of $Y|Z,D$ is a mixture of potential outcomes from principal   strata. Let $f_A(y)$ be the density of outcome for a patient who received $D=A$ when assigned to $Z=A$ (i.e., $f_A(y)=f(y|D=A,Z=A$). Then
\begin{align*}
f_A(y)= \gamma_A^{S2  } f_A^{S2  }(y) +\gamma_A^{S4}  f_A^{S4} (y)+\gamma_A^{S6}   f_A^{S6} (y),
\end{align*}
where $\gamma_A^{S.}=\frac{\pi^{S.}}{\pi^{S2  }+\pi^{S4} +\pi^{S6} }$ and $f_A^{S.}(y)$ is the density function of the outcome for subjects in principal   stratum $S.$ with $D=A$.  
The density $f_A^{S2  }(y)$ and $f_A^{S6} (y)$ can be written as
\begin{align*}
f_A^{S2  }(y) &= \frac{p(D(B)=A|D(A)=A,Y(A)=y) f_A(y)}{p(D(B)=A|D(A)=A)} \\
f_A^{S6} (y) &= \frac{p(D(B)=C|D(A)=A,Y(A)=y) f_A(y)}{p(D(B)=C|D(A)=A)}
\end{align*}
Thus, {\small{
\begin{align}
f_A^{S4}  (y)= \frac{ [1-p(D(B)=C|D(A)=A,Y(A)=y)-p(D(B)=A|D(A)=A,Y(A)=y)]f_A(y)  }{\gamma_A^{S4} }
\label{eq:fm}
\end{align} }}
Similarly, let $f_B(y)$ be the density of outcome for a patient who received $D=B$ when assigned to $Z=B$. Then
\begin{align*}
f_B(y)= \gamma_B^{S1}  f_B^{S1} (y) +\gamma_B^{S4}  f_A^{S4} (y)+\gamma_B^{S5}   f_B^{S5} (y),
\end{align*}
where $\gamma_B^{S.}=\frac{\pi^{S.}}{\pi^{S1} +\pi^{S4} +\pi^{S5} }$ and $f_B^{S.}(y)$ is the density function of the outcome for subjects  in principal   stratum $S.$ with $D=B$. Also, 
\begin{align*}
f_B^{S1} (y) &= \frac{p(D(A)=B|D(B)=B,Y(A)=y) f_A(y)}{p(D(A)=B|D(B)=B)}   \\
f_B^{S5} (y) &= \frac{p(D(A)=C|D(B)=B,Y(B)=y) f_B(y)}{p(D(A)=C|D(B)=B)}
\end{align*}
Thus, {\small{
\begin{align}
f_B^{S4}  (y)= \frac{ [1-p(D(A)=C|D(B)=B,Y(B)=y)-p(D(A)=B|D(B)=B,Y(B)=y)]f_B(y)  }{\gamma_B^{S4} }.
\label{eq:fs} 
\end{align} }}

The mixture proportions $\gamma_A^{S.}$ and $\gamma_B^{S.}$ are  unknown and need to be estimated using the observed data. However, the strata proportions $\pi^{S.}$ are not identifiable using the proportions in the observable $(D,Z)$ strata, because
\begin{align}
p(D=A|Z=A) &= \pi^{S2  }+\pi^{S4} +\pi^{S6}  \nonumber \\
p(D=A|Z=B) &= \pi^{S2  }\nonumber\\
p(D=B|Z=A) &= \pi^{S1} \nonumber\\
p(D=B|Z=B) &= \pi^{S1} +\pi^{S4} +\pi^{S5} \nonumber \\
p(D=C|Z=B) &= \pi^{S3  }+\pi^{S6} \nonumber\\
p(D=C|Z=A) &= \pi^{S3  }+\pi^{S5},
\label{eq:psprobs}
\end{align}
does not have a unique solution.  Also, the density functions  $f_A^{S4} (y)$ and $f_B^{S4}  (y)$ are not  identifiable using the observed data because $p(D(B)=A|D(A)=A,Y(A)=y)$, $p(D(B)=C|D(A)=A,Y(A)=y)$, $p(D(A)=B|D(B)=B,Y(B)=y)$ and $p(D(A)=C|D(B)=B,Y(B)=y)$ are unknown.  So we assume  logit models for these probabilities  and perform the sensitivity analysis based on these models;
\begin{align}
p(D(B)=B|D(A)=A,Y(A)=y) &= w_A(y; \alpha_{0A},\alpha_{1A})= \frac{e^{\alpha_{0A}+\alpha_{1A}y}}{1+e^{\alpha_{0A}+\alpha_{1A}y}} \label{eq:selprobs1} \\
p(D(A)=A|D(B)=B,Y(B)=y) &= w_B(y; \alpha_{0B},\alpha_{1B})= \frac{e^{\alpha_{0B}+\alpha_{1B}y}}{1+e^{\alpha_{0B}+\alpha_{1B}y}}.
\label{eq:selprobs2}
\end{align}
Note that for any fixed ($\alpha_{1A}, \alpha_{1B}$), ($\alpha_{0A}, \alpha_{0B}$) can be determined since $\int f_A^{S4} (y) dy=1$ and $\int f_B^{S4} (y) dy=1$. Therefore, the parameter of interest $\theta$ can be estimated using the four  dimensional sensitivity parameter ($\gamma_A^{S4} ,\gamma_B^{S4} ,\alpha_{1A}, \alpha_{1B})$.

\textbf{Remark.} If it is plausible to assume that at least one of the principal strata $\pi^{S_3}$, $\pi^{S_5}$ or $\pi^{S_6}$ is empty, we can reduce the dimension of the sensitivity parameters to two by estimating $\gamma_A^{S4}$ and $\gamma_B^{S4}$ using the observed data.



\subsection{No Selection Bias}
 Let $R$ be a binary variable which is 1 if a patient is assigned to one of the treatments of interest and 0 otherwise.   Under either of the following two conditions,  the treatment effect  is identifiable when we have access to the entire data (i.e., data includes patients who have received treatment C as well as A and B):

\begin{itemize}
\item[A.1] $p(R=0|Z, U,\bX) =p(R=0| U, \bX)$ where $\bX$ and $U$ are measured and unmeasured confounders, respectively.  
\item[A.2] $p(R=0|Z, U,\bX) =p(R=0| Z,\bX)$.
\end{itemize}

\begin{figure}
\centerline{
\xymatrix{
& X \ar[d] \ar[rd] \ar[rrd] & U \ar[rd] \ar[d] \ar[ld] \\
Z  \ar[r]   \ar@/^1pc/[rr] &R \ar[r] &D \ar[r] & Y\\
  } \hspace{.8cm}
\xymatrix{
& X \ar[d] \ar[rd] \ar[rrd] & U \ar[rd] \ar[d] \ar[ld] \\
Z     \ar@/^1pc/[rr] &R \ar[r] &D \ar[r] & Y\\
  }
   \hspace{.8cm}
\xymatrix{
& X \ar[d] \ar[rd] \ar[rrd] & U \ar[rd] \ar[d]  \\
Z  \ar[r]   \ar@/^1pc/[rr] &R \ar[r] &D \ar[r] & Y\\
  }
  }
   \caption{Causal diagram. Conditioning on subjects with $R=1$ leads to a bias treatment effect estimate in the left figure. Assumptions A.1 and A.2 hold for the center and right figure, respectively.  } 
   \label{fig:dag}
\end{figure}
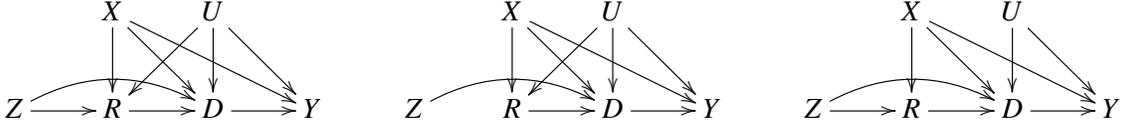

Assumption A.1 implies that the IV is unrelated to the patient's selection. In other words, a patient would have the same chance of being selected for both values of the IV.  A.2 means that the selection probability is independent of the unmeasured confounders given the instrumental variable and measured confounders. Figure \ref{fig:dag} depicts the association between different variables. The left plot presents a scenario in which analyzing the preselected data will cause bias because conditioning on $R$ will open the path $Z\rightarrow R \rightarrow U \rightarrow Y$ which violates one of the IV assumptions. This path is closed under either of assumptions A.1 or A.2 as shown in the center and right  plots, respectively. 

Under either of assumptions A.1 or A.2, the treatment effect can be identified. Specifically, under assumption A.1, the principal   strata $S5$ and $S6$ does not exist. Thus
\begin{align*}
\E[Y|Z=A,D=A]&=\gamma^{S2  } \E[Y^A|PS=S2  ]+\gamma^{S4} \E[Y^A | PS=S4]\\
\E[Y|Z=B,D=B]&=\gamma^{S1}  \E[Y^B|PS=S1]+\gamma^{S4} \E[Y^B | PS=S4]\\
\E[Y|Z=A,D=B]&=\gamma^{S1}  \E[Y^B|PS=S1]\\
\E[Y|Z=B,D=A]&=\gamma^{S2  } \E[Y^A|PS=S2  ].
\end{align*}
These equations can be used to identify $\theta=\E[Y^A-Y^B|PS=S4]$. Also, under assumption A.2, $\theta$ can be estimated using the following three-step procedure:
\begin{itemize}
\item[Step1.] Using the entire data, estimate $ \varpi=p(R=1|\bX,Z)$ by postulating a logit model on $S$. 
\item[Step2.] Include patients who have been assigned to either treatments A or B.
\item[Step3.] Estimate the treatment effect $\theta$ using 2SLS IV approach.  At each stage fit a weighted least square where weights are $\hat \varpi$. 
\end{itemize}

Fitting weighted least squares accounts for the selection procedure and hence avoids the bias by creating a pseudo-population in which there is no association between $(\bX,Z)$ and $R$.

\subsection{Inference Procedure}

Let $\bY_A=(Y_{A1}, Y_{A2},...,Y_{An_A})$ and $\bY_B=(Y_{B1}, Y_{B2},...,Y_{Bn_B})$ denote the outcome of patients whose received treatment matches their IV value for $Z=A$ and $Z=B$, respectively. We assume that samples are independent within and between each IV group and  identically distributed within each group. 


For any fixed $(\gamma_A^{S4} ,\gamma_B^{S4} ,\alpha_{1A}, \alpha_{1B} )$, $(\hat \alpha_{0A}, \hat \alpha_{0B})$ is defined as a solution to 
\begin{align*}
\int_{-\infty}^{-\infty} \frac{e^{\alpha_{0A}+\alpha_{1A}y}}{1+e^{\alpha_{0A}+\alpha_{1A}y}} d\hat F_A(y) &= \gamma_A^{S4} \\
\int_{-\infty}^{-\infty} \frac{e^{\alpha_{0B}+\alpha_{1B}y}}{1+e^{\alpha_{0B}+\alpha_{1B}y}} d\hat F_B(y) &= \gamma_B^{S4} 
\end{align*}
where $\hat F_A(y)$ and $\hat F_B(y)$ are empirical distributions calculated from the observed samples $\bY_A$ and $\bY_B$. These equations can be solved using a one dimensional grid search on real line. 

We estimate the treatment effect $\theta$ by
\[
\hat \theta = \left[\frac{1}{\gamma_A^{S4} n_A} \sum_{j=1}^{n_A} y_{Aj}w_A(y_{Aj} ; \hat \alpha_{0A},\alpha_{1A})  \right]   -   \left[  \frac{1}{\gamma_B^{S4} n_B} \sum_{j=1}^{n_B}   y_{Bj}w_B(y_{Bj} ; \hat \alpha_{0B},\alpha_{1B})  \right].
\]
where $w_A(.)$ and $w_B(.)$ are defined in  (\ref{eq:selprobs1}) and (\ref{eq:selprobs2}). By bootstrap sampling from  $\bY_A$ and $\bY_B$, we can estimate the standard error of $\hat \theta$ as the standard error of the estimated treatment effect from each bootstrapped sample. 

\section{Simulation Studies} \label{sec:simulation}

\begin{figure}[t]
\centering
\makebox{\includegraphics[scale=.40]{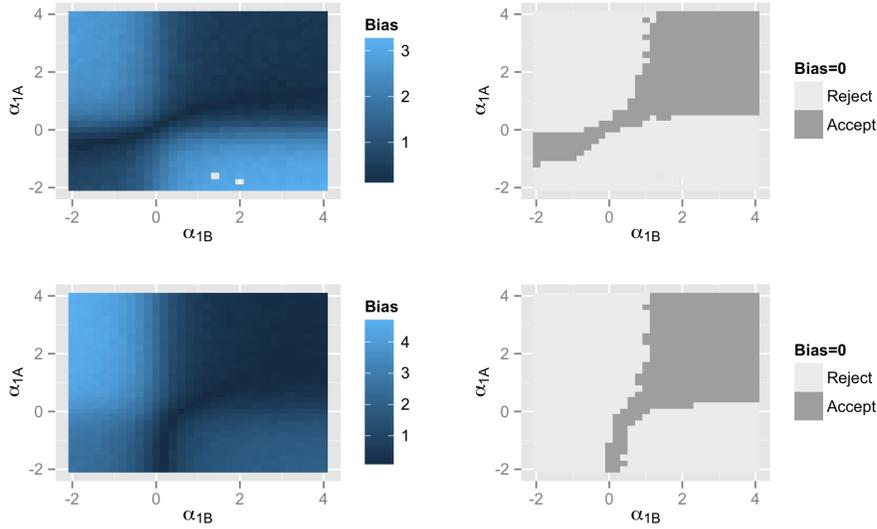}}
\caption{ { Simulation study: Sensitivity analysis of $\theta=0.80$ when $\gamma_A^{S4}$ and $\gamma_B^{S4}$ are known. The true sensitivity parameter values are $(\alpha_{1A},\alpha_{1B})=(1,2)$. The darker area in left panel indicates the smaller bias and the darker area in the right panel shows the region in which the bias is not significant at $5\%$.  The first and second rows represent Scenario 1 and 2, respectively.  }}
\label{fig:2dimsens}
\end{figure}

Through simulations we examine the performance of our proposed sensitivity analysis assuming known and unknown $(\gamma_A^{S4} ,\gamma_B^{S4})$. We assume two scenarios with the following principal   strata proportions: Scenario 1: 
\begin{itemize}
\item Scenario 1: $\pi^{S1} =\pi^{S2  }=\pi^{S3  }=\pi^{S5} =0.1$, $\pi^{S4} =0.3$ and $\pi^{S6} =0.3$ which gives us $(\gamma_A^{S4} ,\gamma_B^{S4} )=(\frac{0.3}{0.7}, \frac{0.3}{0.5})$.
\item Scenario 2: $\pi^{S1} =\pi^{S2  }=\pi^{S3  }=0.1$, $\pi^{S4 }=0.3$, $\pi^{S5} =0.4$  and $\pi^{S6} =0.0$ which gives us $(\gamma_A^{S4 },\gamma_B^{S4 })=(\frac{0.2}{0.3}, \frac{0.2}{0.7})$.

\end{itemize}
The true amount of selection bias is determined by the parameters $\alpha_{1A}$ and $\alpha_{1B}$ in (\ref{eq:selprobs1}) and (\ref{eq:selprobs2}), respectively. The outcome of individuals with $D=A$ and $Z=A$ are generated from a normal distribution with mean 2.5 and standard error 2 (i.e., $f_A(y)=N(2.5,2)$). Also, the outcome of individuals with $D=B$ and $Z=B$ are generated from a normal distribution with mean $\mu$ and standard error 2 (i.e., $f_B(y)=N(\mu,2)$). For each set of $(\alpha_{1A},\alpha_{1B})$, the parameter $\mu$ is determined such that the true treatment effect among compliers is $\theta=0.00$, 0.50 and 0.80. We generate samples using rejection sampling from densities $f_B^{S4 } (y)$ and $f_B^{S4 } (y)$ according to (\ref{eq:fm}) and (\ref{eq:fs}), respectively.

Table \ref{tab:senspow} presents the statistical power of detecting the true treatment effect $\theta=0.0$, 0.50, and 0.80 based on 500 datasets of sizes n=500  with $(\alpha_{1A},\alpha_{1B})=(1,2)$, (1,1) and (0,0). The nominal type-I error rate is 5\%.  For both scenarios when the true amount of selection bias is presumed, the empirical and nominal sizes are close. In general, the empirical sizes are inflated when the selection bias parameters are not set correctly. For some of the combinations of the true and presumed sensitivity parameter values the power and type-I error rate are much higher than  the nominal one which is caused by the overestimation of the treatment effect. Also, in scenario 2, when there is no selection bias  (i.e., true $(\alpha_{1A},\alpha_{1B})=(0,0)$) and we presumed $(\alpha_{1A},\alpha_{1B})=(1,1)$ the power is very low because of the underestimation of the treatment effect.  Table \ref{tab:sensest} shows the bias ($0.80-\hat \theta$), standard error and mean squared error of the treatment effect estimation for different combinations of the true and presumed sensitivity parameters when the true treatment effect $\theta=0.80$. This table highlights the over- and underestimation phenomenon discussed earlier. Specifically, when the true sensitivity parameters are $(\alpha_{1A},\alpha_{1B})=(1,2)$, presuming $(\alpha_{1A},\alpha_{1B})=(0,2)$ will result in underestimating the counterfactual mean outcome $\E[Y^A | PS=S4]$. This is because when we set the presumed $\alpha_{1A}$ less than the true value, the larger outcomes are less likely to be a complier among patients with $D=Z=A$. In other words it will shift the distribution of this compliers to left. Similarly when the presumed $\alpha_{1A}$ more than the true value, $\E[Y^A | PS=S4]$ will be overestimated (See rows 3 and 4 in Table \ref{tab:sensest}).

Figure \ref{fig:2dimsens} shows how the treatment effect estimate changes by the sensitivity parameters when the true $(\alpha_{1A},\alpha_{1B})=(1,2)$ and $\theta=0.80$. The darker area in left panel of Figure \ref{fig:2dimsens} indicates the smaller bias and the darker area in the right panel shows the region in which the bias is not significant at $5\%$ (i.e., the confidence interval for $(0.80-\hat \theta)$ contains zero). See also Table \ref{tab:sensest}.

\begin{figure}[t]
\centering
\makebox{\includegraphics[scale=.52]{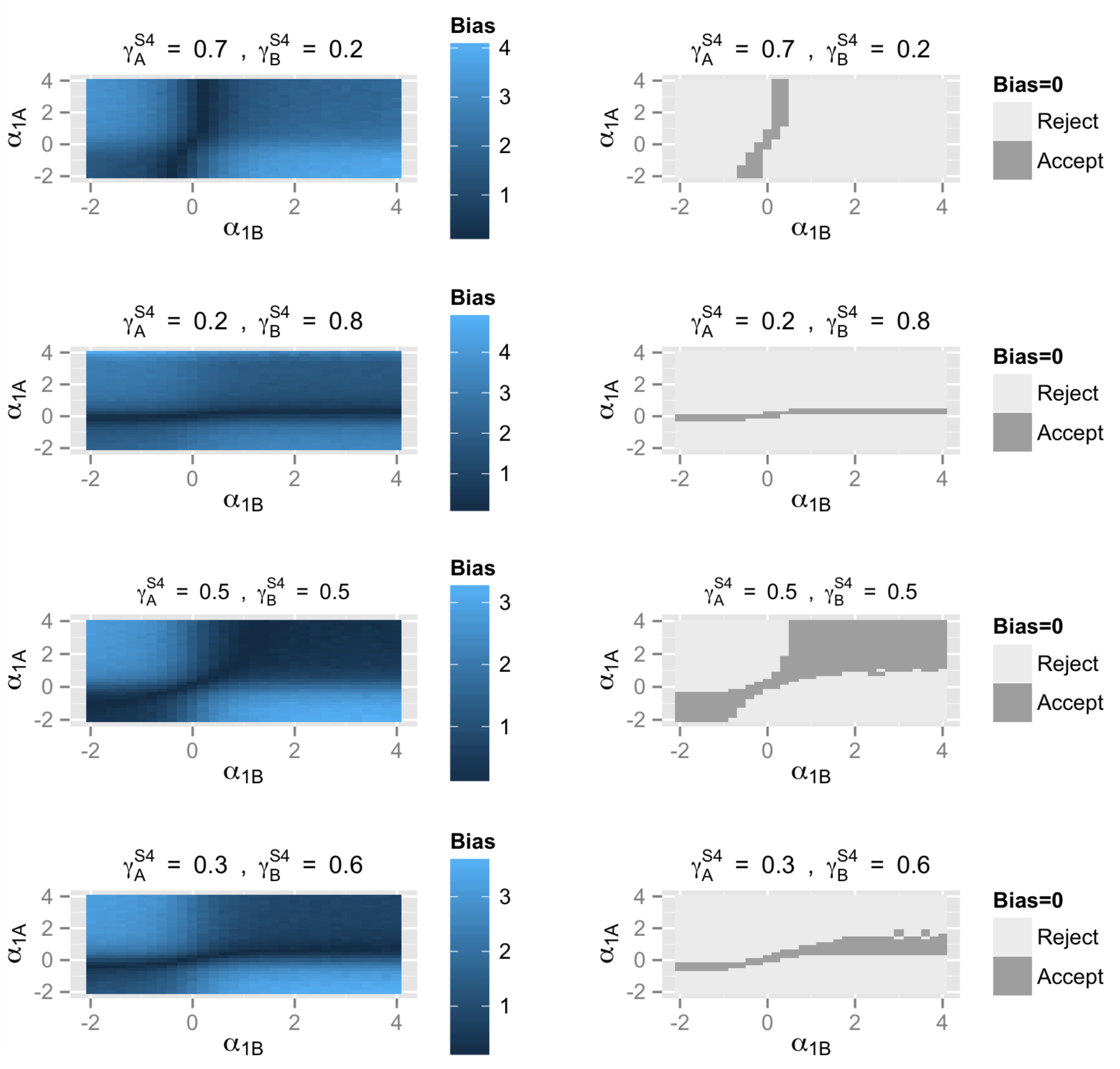}}
\caption{ {  Simulation study (Scenario 1): Sensitivity analysis of $\theta=0.80$. Sensitivity parameters are $(\gamma_A^{S4} ,\gamma_B^{S4} ,\alpha_{1A}, \alpha_{1B} )$. The darker area in left panel indicates the smaller bias and the darker area in the right panel shows the region in which the bias is not significant at $5\%$. The true vector of sensitivity parameters are $(\gamma_A^{S4} ,\gamma_B^{S4} ,\alpha_{1A}, \alpha_{1B} )=\left(\frac{3}{7},\frac{3}{5},1,2\right)$. }}
\label{fig:4dimsens}
\end{figure}

\begin{figure}[t]
\centering
\makebox{\includegraphics[scale=.52]{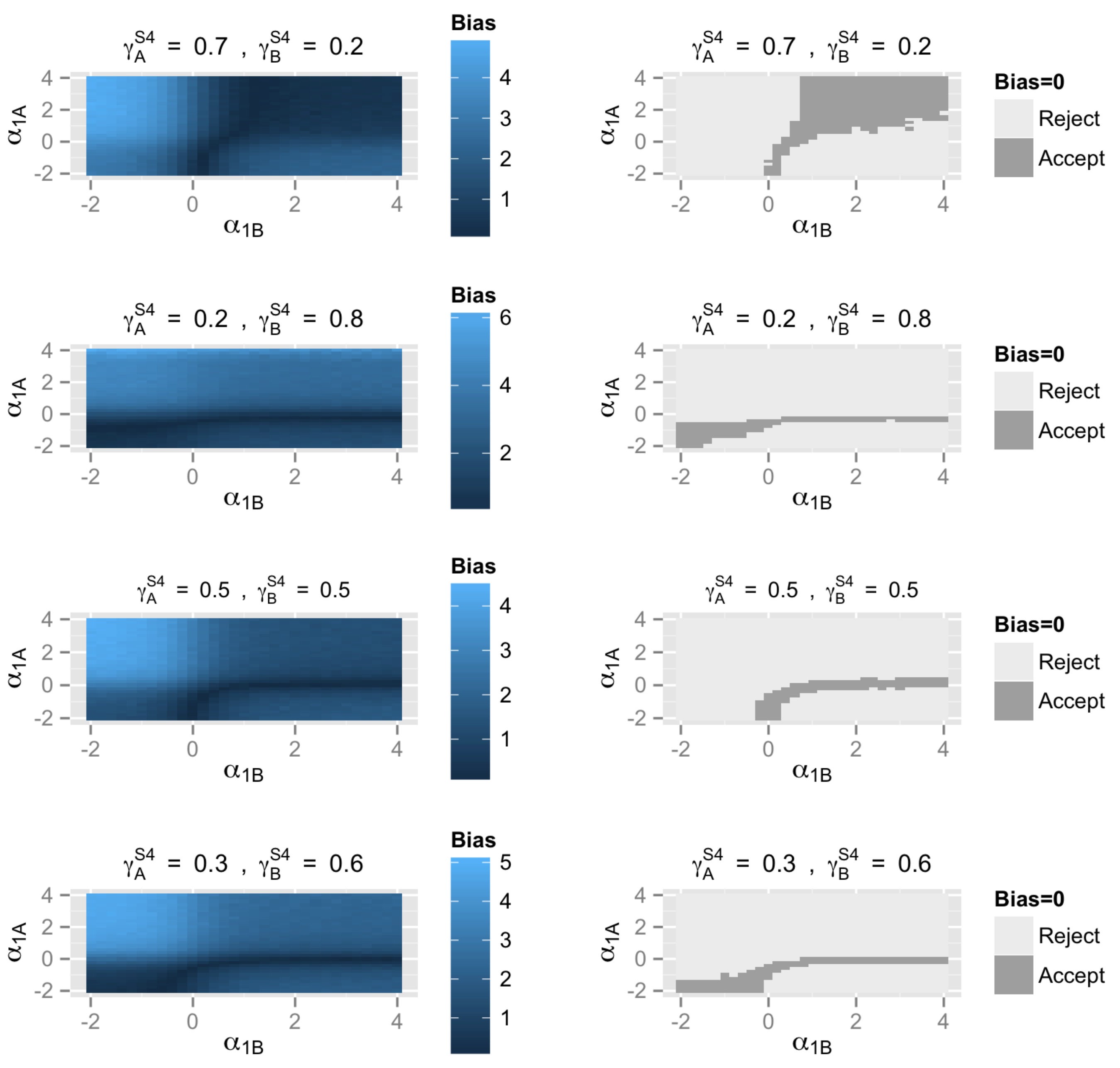}}
\caption{ {  Simulation study (Scenario 2): Sensitivity analysis of $\theta=0.80$. Sensitivity parameters are $(\gamma_A^{S4} ,\gamma_B^{S4} ,\alpha_{1A}, \alpha_{1B} )$. The darker area in left panel indicates the smaller bias and the darker area in the right panel shows the region in which the bias is not significant at $5\%$. The true vector of sensitivity parameters are $(\gamma_A^{S4} ,\gamma_B^{S4} ,\alpha_{1A}, \alpha_{1B} )=\left(\frac{2}{3},\frac{2}{7},1,2 \right)$. }}
\label{fig:s24dimsens}
\end{figure}

\begin{table}[t]
\caption{\label{tab:senspow}  Simulation study: Power analysis of $\theta=0.00, 0.50$ and 0.80 when $\gamma_A^{S4}$ and $\gamma_B^{S4}$ are known.  } \centering {
\begin{tabular}{*{5}{c} |*{3}{c} } \hline
  && \multicolumn{3}{c}{Scenario 1} & \multicolumn{3}{c}{Scenario 2}\\
True $(\alpha_{1A},\alpha_{1B})$ &Presumed $(\alpha_{1A},\alpha_{1B})$& 0.00 &  0.50 & 0.80& 0.00 & 0.50 & 0.80\\ \hline
\textbf{(1,2)} & \textbf{(1,2)} & 0.04&0.60& 0.91&0.04 &0.52 &0.86 \\
(1,2) & (1,1)  & 0.12&0.81 &0.98 &0.22 &0.88 &0.97 \\
(1,2) & (0,0) & 0.16 &0.36 &0.78&0.99 &1.00& 1.00 \\ \hline
\textbf{(1,1)} & \textbf{(1,1)}& 0.05 &0.58 &0.91 &0.05 &0.48 &0.85\\
(1,1) & (2,2) & 0.08 &0.71 &0.95&0.10 &0.24 &0.61\\
(1,1) & (0,0) & 0.35 &0.05 &0.40&0.90 &1.00 &1.00\\ \hline
\textbf{(0,0)} & \textbf{(0,0)} & 0.04 &0.61 &0.92&0.05 &0.60 &0.94\\
(0,0) & (1,0) & 1.00 &1.00 &1.00&0.80 &1.00 &1.00\\
(0,0) & (1,1) & 0.34 &0.96 &1.00&0.81 &0.00 &0.03\\
 \hline
\end{tabular}}
\end{table}

\begin{table}[t]
\caption{\label{tab:sensest}  Simulation study: Sensitivity analysis of $\theta=0.80$ when $\gamma_A^{S4}$ and $\gamma_B^{S4}$ are known. Bias$=0.80-\hat \theta$.  } \centering {
\begin{tabular}{*{5}{c} |*{3}{c} } \hline
  && \multicolumn{3}{c}{Scenario 1} & \multicolumn{3}{c}{Scenario 2}\\
True $(\alpha_{1A},\alpha_{1B})$ &Presumed $(\alpha_{1A},\alpha_{1B})$& Bias &  S.D. & MSE& Bias & S.D. & MSE\\ \hline
(1,2) & (1,2) &-0.01 &0.28 &0.08 & 0.01&0.30& 0.09\\
(1,2) & (1,1)  &-0.22 &0.29 &0.14& -0.38&0.32 &0.25 \\
(1,2) & (0,2)  &1.38 &0.25 &1.97& 0.81&0.29 &0.74 \\
(1,2) & (1,0) &-1.18 &0.25& 1.43& -2.17 &0.28 &4.82 \\
(1,2) & (0,0) &0.21 &0.23 &0.10& -1.34 &0.25 &1.86\\ \hline
(1,1) & (1,1) &0.00 &0.28 &0.08 & -0.01&0.30& 0.10\\
(1,1) & (1,2)  &0.19 &0.28 &0.11& 0.39&0.31 &0.26 \\
(1,1) & (0,2)  &1.58 &0.27 &2.58& 1.17&0.30 &1.49 \\
(1,1) & (1,0) &-0.98 &0.26& 1.02& -1.77 &0.28 &3.20 \\
(1,1) & (0,0) &0.41 &0.23 &0.23& -0.95 &0.27 &0.98\\ \hline
\end{tabular}}
\end{table}

\begin{table}[t]
\caption{\label{tab:4dimsensest}  Simulation study: Power analysis of $\theta=0.00, 0.50$ and 0.80.  The true vector of sensitivity parameters in scenario 1 and 2 are $(\gamma_A^{S4} ,\gamma_B^{S4} ,\alpha_{1A}, \alpha_{1B} )=(\frac{3}{7},\frac{3}{5},1,2)$, and $(\frac{2}{3},\frac{2}{7},1,2)$, respectively.  } \centering {
\begin{tabular}{*{4}{c} |*{3}{c} } \hline
 Presumed & \multicolumn{3}{c}{Scenario 1} & \multicolumn{3}{c}{Scenario 2}\\
   $(\gamma_A^{S4} ,\gamma_B^{S4}, \alpha_{1A},\alpha_{1B})$& 0.00 &  0.50 & 0.80& 0.00 & 0.50 & 0.80\\ \hline
 (0.3,0.6,1,2) &0.38 &0.90 &0.98 & 1.00&1.00& 1.00\\
 (0.3,0.6,0,2)  &1.00 &0.00 &0.00& 0.07&0.73 &0.96 \\
 (0.3,0.6,1,0) &1.00 &1.00& 1.00& 1.00 &1.00 &1.00 \\
 (0.3,0.6,0,0) &0.12 &0.35 &0.80& 1.00 &1.00 &1.00\\ \hline
 (0.5,0.5,1,2) &0.40 &0.08 &0.38 & 0.93&1.00& 1.00\\
 (0.5,0.5,0,2)  &1.00 &0.00 &0.00& 0.04&0.40 &0.75 \\
 (0.5,0.5,1,0) &0.96 &1.00& 1.00& 1.00 &1.00 &1.00 \\
 (0.5,0.5,0,0) &0.10 &0.29 &0.76& 1.00 &1.00 &1.00\\ \hline
 (0.2,0.8,1,2) &0.98 &1.00 &1.00 & 1.00&1.00& 1.00\\
 (0.2,0.8,0,2)  &0.94 &0.00 &0.05& 0.61&1.00 &1.00 \\
 (0.2,0.8,1,0) &1.00 &1.00& 1.00& 1.00 &1.00 &1.00 \\
 (0.2,0.8,0,0) &0.06 &0.24 &0.65& 1.00 &1.00 &1.00\\ \hline
(0.7,0.2,1,2) &1.00 &0.00 &0.00 & 0.30&0.08& 0.32\\
 (0.7,0.2,0,2)  &1.00 &0.00 &0.00& 0.98&0.00 &0.00 \\
 (0.7,0.2,1,0) &0.65 &0.99& 1.00& 1.00 &1.00 &1.00 \\
 (0.7,0.2,0,0) &0.12 &0.39 &0.87& 1.00 &1.00 &1.00\\ \hline
\end{tabular}}
\end{table}

\subsection{Unknown $(\gamma_A^{S4} ,\gamma_B^{S4})$}

 In this case our sensitivity parameter space is $(\gamma_A^{S4} ,\gamma_B^{S4} ,\alpha_{1A}, \alpha_{1B} )$. Figures \ref{fig:4dimsens} and \ref{fig:s24dimsens} present the bias in the treatment effect estimation for different values of the presumed sensitivity parameters based on scenarios 1 and 2, respectively. In these figures the true $(\gamma_A^{S4} ,\gamma_B^{S4} ,\alpha_{1A},\alpha_{1B})=(\frac{3}{7},\frac{3}{5},1,2)$ and $(\gamma_A^{S4} ,\gamma_B^{S4} ,\alpha_{1A},\alpha_{1B})=(\frac{2}{3},\frac{2}{7},1,2)$ in scenarios 1 and 2, respectively. Also the true treatment effect is  $\theta=0.80$.  The darker area in the left panel  shows the region in the sensitivity parameter space in which the bias is smaller and right panel shows the area in which the bias is not significant based on 500 samples (i.e., the confidence interval for $(0.80-\hat \theta)$ contains zero).  The shape of the area which leads to an unbiased  estimate changes by the value of the presumed $(\gamma_A^{S4} ,\gamma_B^{S4} )$ and, in general, when the presumed values of $(\gamma_A^{S4},\gamma_B^{S4} )$  are far from the true values, the unbiased area is small which means it would be less likely to find an unbiased treatment effect estimate using sensitivity analysis (see also Table \ref{tab:4dimsensest2} in Appendix A).  We have tabulated the power and type-I error rate for different values of the presumed sensitivity parameters in Table \ref{tab:4dimsensest}. This Table summarizes the result for three different values of the true treatment effect $\theta=0.00, 0.50$, and 0.80. The nominal type-I error rate is 5\%.

\section{Application to Real Data } \label{sec:application}

The Health Improvement Network (THIN) is a large database, which contains the electronic medical record of more than 11 million patients. The data contains longitudinal measurements of diagnostic and prescription data and baseline characteristics collected from over 500 general practices in the UK. We are interested in assessing  the effect of metformin and sulfonylureas on body mass index (BMI, calculated as mass in kilograms divided by the square of height in meters) among  diabetic patients. Our focus is on patients who are  taking these treatments as an initial treatment and the BMI is measured two years after treatment initiation. In our analysis we coded metformin and sulfonylureas  as $A$ and $B$, respectively.

In the modern era, metformin is universally acknowledged as the appropriate first-line medication for diabetes type 2 except where contraindicated \citep{nathan2009american}.  While this dominance was being established in the late 1990s and early 2000s, sulfonylureas were the competing first line agent. During that time the prevalence of metformin use in clinical practice rose very quickly, at the expense of sulfonylureas.  

The dynamic aspect of the preference justifies the use of provider preference as an IV. We define the preference as a time-varying quantity such that for each practice ID, the IV is defined as an average of metformin use during each two year timeframe from 1998 to 2012. We assign $Z=A$ if the average is more than 50\% and $B$ otherwise. Thus, a particular practice ID may have $Z=A$ for one period of time and $B$ for the other.

\begin{figure}[t]
\centering
\makebox{\includegraphics[scale=.38]{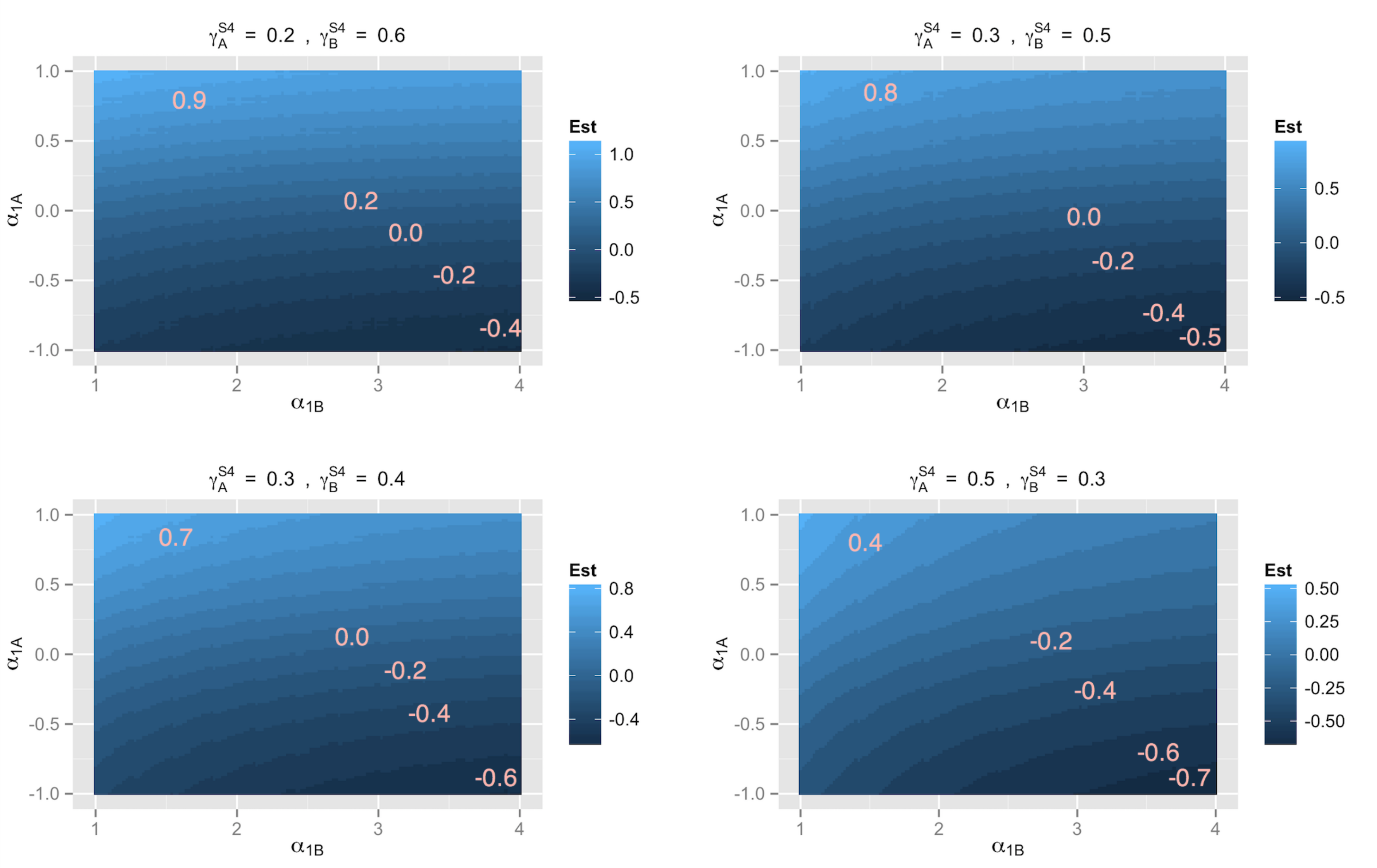}}
\caption{ {  THIN data: Sensitivity analysis for comparing the effect of metformin with sulfonylureas on BMI. Numbers on the plot present  treatment effect estimates. }}
\label{fig:thin4dimsens}
\end{figure}

Our data includes patients who have been assigned to medications other than metformin or sulfonylureas as an initial therapy.  Figure \ref{fig:thin4dimsens} presents the sensitivity analysis for the estimation of the treatment effect of interest based on a four dimensional sensitivity parameter $(\gamma_A^{S4} ,\gamma_B^{S4} ,\alpha_{1A},\alpha_{1B})$.  The parameter $\alpha_{1A}$ identifies the probability that a patient with a particular value of BMI would have taken $D=B$ if $Z=B$ (i.e., seeing a physician with sulfonylureas preference) given that she has taken $D=A$ with $Z=A$ (i.e., seeing a physician with metformin preference). The parameter $\alpha_{1B}$ has a similar interpretation and the parameter $\gamma_A^{S4} $ ($\gamma_B^{S4} $) is defined as a chance of being a complier for patients who have been assigned to $Z=A(B)$ and have taken $D=A(B)$. Since the association of sulfonylureas with weight gain is known it makes sense to assume that the true value of the parameter $\alpha_{1A}$ should be negative and $\alpha_{1B}$ should be positive. Also on average patients who have taken sulfonylureas ($D=B$) given $Z=B$ are more likely to be compliers compare with those who have taken metformin ($D=A$) given $Z=A$. This means that it is very likely that  $\gamma_A^{S4} < \gamma_B^{S4} $.

The range of the treatment effect estimate varies by different choices of $\gamma_A^{S4} $ and $\gamma_B^{S4} $. Specifically, for   $\gamma_A^{S4}=0.2 $ and $\gamma_B^{S4}=0.6 $, the treatment effect lies in $[-0.5,1]$; while for $\gamma_A^{S4}=0.5 $ and $\gamma_B^{S4}=0.3 $, it lies in $[-0.7,0.5]$. 
Based on our sensitivity analysis, it is unlikely that metformin be associated with increase in BMI and a large negative  effect (i.e., $\E[Y^A-Y^B|PS=S4]$) is also unlikely because it requires very small and large values of $\alpha_{1A}$ and $\alpha_{1B}$, respectively. Hence, if anything, the treatment effect of interest should lie in $[-0.6,-0.1]$. 

\section{Discussion} \label{sec:conclude}

This manuscript shows that IV analysis methods may fail to provide an unbiased estimate of treatment effects when analyzing data in which subjects are selected based on their received treatment. Specifically, this selection bias happens if  the chance of including patients in the preselected data varies based on  their IV value and/or some unmeasured confounders.   For example, suppose we are interested in studying the comparative effect of treatment $A$ and $B$ while there is a third choice of treatment $C$. Suppose patients with more severe stages of the disease are more likely to be treated with $C$ if seeing a doctor with $A$ preference. Then if we analyzed a data that only includes patients who have received treatment $A$ or $B$, severe cases will be under represented among patients who have seen doctors with $A$ preference. This means that in the preselected data, the IV is associated with an unmeasured confounder which makes the IV invalid and results in a biased treatment effect estimate.

Within the principal strata framework, we develop a sensitivity analysis that can be used to estimate the treatment effect among compliers as a function of a vector of sensitivity parameters. Specifically, the sensitivity parameters are used to identify the probability of being a complier given the IV value, received treatment and the observed outcome. The dimension of the sensitivity parameter can be reduced if it is plausible to assume that there is no always  $C$ takers.  This is because the proportions in the principal strata are identifiable which means $\gamma_A^{S4} $ and $\gamma_B^{S4} $ can be estimated using the observed data.

\section*{Acknowledgment}

This work is partially supported by NSF grant SES-1260782. 

\appendix
\section*{Appendix}
\label{app:yopt}

In this Appendix, we present the supplementary materials.

\section*{Appendix: A}

Table \ref{tab:4dimsensest2} shows the bias ($0.80-\hat \theta$), standard error and mean squared error of the treatment effect estimation for different combinations of the true and presumed sensitivity parameters when the true treatment effect $\theta=0.80$. 

\begin{table}[t]
\caption{\label{tab:4dimsensest2}  Simulation study: Sensitivity analysis of $\theta=0.80$. The true vector of sensitivity parameters in scenario 1 and 2 are $(\gamma_A^{S4} ,\gamma_B^{S4} ,\alpha_{1A}, \alpha_{1B} )=(\frac{3}{7},\frac{3}{5},1,2)$, and $(\frac{2}{3},\frac{2}{7},1,2)$, respectively. } \centering {
\begin{tabular}{*{4}{c} |*{3}{c} } \hline
  & \multicolumn{3}{c}{Scenario 1} & \multicolumn{3}{c}{Scenario 2}\\
Presumed $(\gamma_A^{S4} ,\gamma_B^{S4} ,\alpha_{1A},\alpha_{1B})$& Bias &  S.D. & MSE& Bias & S.D. & MSE\\ \hline
 (0.3,0.6,1,2) &-0.37 &0.29 &0.22 & -1.92&0.34& 3.79\\
 (0.3,0.6,0,2)  &1.37 &0.24 &1.95& -0.18&0.30 &0.12 \\
 (0.3,0.6,1,0) &-1.54 &0.27& 2.45& -3.07 &0.34 &9.55 \\
 (0.3,0.6,0,0) &0.23 &0.24 &0.11& -1.32 &0.27 &1.83\\ \hline
 (0.5,0.5,1,2) &0.46 &0.26 &0.28 & -1.09&0.29& 1.28\\
 (0.5,0.5,0,2)  &1.65 &0.26 &2.80& 0.11&0.27 &0.08 \\
 (0.5,0.5,1,0) &-0.95 &0.27& 1.04& -2.57 &0.28 &6.72 \\
 (0.5,0.5,0,0) &0.20 &0.23 &0.09& -1.35 &0.26 &1.89\\ \hline
 (0.2,0.8,1,2) &-1.22 &0.31 &1.59 & -2.76&0.39& 7.80\\
 (0.2,0.8,0,2)  &0.87 &0.25 &0.81& -0.69&0.27 &0.55 \\
 (0.2,0.8,1,0) &-1.86 &0.29& 3.54& -3.42 &0.36 &11.88 \\
 (0.2,0.8,0,0) &0.23 &0.23 &0.11& -1.32 &0.25 &1.79\\ \hline
(0.7,0.2,1,2) &1.98 &0.34 &4.03 & 0.44&0.32& 0.30\\
 (0.7,0.2,0,2)  &2.72 &0.33 &7.54& 1.20&0.32 &1.53 \\
 (0.7,0.2,1,0) &-0.55 &0.24& 0.36& -2.09 &0.27 &4.45 \\
 (0.7,0.2,0,0) &0.18 &0.25 &0.10& -1.34 &0.26 &1.88\\ \hline
\end{tabular}}
\end{table}

\pagebreak
\bibliographystyle{DeGruyter}
\bibliography{IV-Bias}

\label{lastpage}
\end{document}